\documentclass[prb,aps,twocolumn]{revtex4} 
\usepackage{psfig}
\newcommand{\be}{\begin{equation}} 
\newcommand{\ee}{\end{equation}}
\newcommand{\e}{\emph} 
\newcommand{\bb}{\textbf}

\def\beginwide{

\noindent
        \rule{3.5in}{.1mm}\rule{.1mm}{5mm}  \widetext{
\medskip }
\def\beginwidetop{
        \end{multicols} \vspace*{-0.5cm} \noindent
        \widetext \medskip }
\def\endwide{
        \hspace*{3.35in}~\rule[-5mm]{.1mm}{5mm}\rule{3.5in}{.1mm}
}
\noindent }
\def\endwidebottom{
        \begin{multicols}{2} \vspace*{-1.0cm} \noindent }

\begin{document}

\title{{\Large Local Rigidity in Sandpile Models}}

\author{S. Ciliberti$^1$, G. Caldarelli$^1$, 
V. Loreto$^{1,2}$ and
L. Pietronero$^{1,2}$} \affiliation
{$^1$INFM UdR Roma1 and ``La Sapienza'' University,
Physics Department, P.le A. Moro 5, 00185 Rome, Italy \\ 
$^2$ Center for Statistical Mechanics
and Complexity (SMC), P.le A. Moro 5, 00185 Rome, Italy}

\begin{abstract}
We address the problem of the role of 
the concept of local rigidity in
the family of sandpile systems.  We define 
rigidity as the ratio
between the critical energy and the amplitude 
of the external
perturbation and we show, in the framework of 
the Dynamically Driven
Renormalization Group (DDRG), that any finite 
value of the rigidity in
a generalized sandpile model renormalizes 
to an infinite value at the
fixed point, i.e. on a large scale.  The 
fixed point value of the
rigidity allows then for a non ambiguous 
distinction between
sandpile-like systems and diffusive systems.  
Numerical simulations
support our analytical results.
\end{abstract}

\maketitle

\section{Introduction}
The concept of Self Organized Criticality
(SOC)~\cite{btw,jen,review_braz} has been proposed 
as a unifying
theoretical framework to describe a vast class of 
driven systems that
evolve ``spontaneously'' to a stationary state 
characterized by power
law distributions of dissipation events.  While 
originally SOC systems
were associated to the absence of tuning parameters, 
it is now
clear~\cite{vz97,review_braz} that most of 
the proposed systems show a
critical behaviour only in the limit of an 
infinite separation of time
scales, i.e. if some suitable parameter (e.g. 
the dissipation and the
driving rate in sandpile models~\cite{vzl96}, 
the temperature in some
growth models~\cite{verg}, etc.) is set to zero. 
These results
suggested a new definition of SOC~\cite{gcp} as 
the theory of
dynamical processes that lead a system to the 
critical steady state if
the critical value of the control parameter is zero.

The stabilizing effect of a threshold~\cite{caf} 
in the class of
sandpile models is discussed by introducing a 
\emph{local rigidity}
defined as the ratio between the critical 
energy and the amplitude of
the external perturbation, 
$r\equiv\varepsilon_c/\delta\varepsilon$.
A finite rigidity allows the system to assume 
a large number of
metastable configurations and this let the 
process show a power law
distribution of avalanches.  In~\cite{caf} 
it has been shown by means
of numerical simulations that in the limit 
$r\rightarrow 0$ the system
becomes a diffusive one characterized by 
only infinite avalanches. In
this paper we show how, in the framework 
of the so-called Dynamically
Driven Renormalization Group (DDRG)~\cite{pvz94,ddrg}, 
a finite
rigidity in the microscopic dynamics is crucial 
in order for the
system to reach spontaneously a critical 
stationary state instead of a
diffusive one.  The basic idea is that at 
large scale the value of the
rigidity does not depend on the microscopic 
value and that this
``coarse grained'' value allows for a non 
ambiguous distinction
between a diffusive and a SOC system.
 
\section{The renormalization of the rigidity}

In Sandpile Models we assign an integer 
variable (``energy'')
$\varepsilon_i$ on each site $i$ of a $d$-dimensional 
hypercubic
lattice. At each time step an energy grain 
is added on a randomly
chosen site, and then the system is allowed 
to relax according to a
particular stability criterion depending 
on a fixed threshold
(e.g. the energy, or the slope -defined 
as the difference between the
energy of two nearest neighbours- exceeding 
a critical value).
Infinite slow driving, i.e. an infinite 
separation of time scales, is
built into the model: during the updating 
process the external input
stops. The original BTW sandpile model~\cite{btw} 
is a critical height
model with $\varepsilon_c=4$ and external 
input $\delta\varepsilon=1$.

We consider a generalized sandpile model 
with critical height
$\varepsilon_c$ and external perturbation 
$\delta\varepsilon$ for the
microscopic dynamics.  
If $\varepsilon_i>\varepsilon_c$ at time $t$
then $\varepsilon_i=0$ at time $t+1$ and 
the $q$ nearest neighbours of
site $i$ increase their value of $\varepsilon_i/q$.  
At the generic
scale these quantities have the values 
$\varepsilon_c^{(k)}$ and
$\delta\varepsilon^{(k)}$. The height 
of the stable sites is less than
$\varepsilon_c^{(k)}-\delta\varepsilon^{(k)}$, 
for the critical sites
the energy lies between 
$\varepsilon_c^{(k)}-\delta\varepsilon^{(k)}$
and $\varepsilon_c^{(k)}$ while for unstable 
sites it is larger than
$\varepsilon_c^{(k)}$.  We study the behaviour of
$\varepsilon_c^{(k)}$ and 
$\delta\varepsilon^{(k)}$ under the
renormalization flow of the DDRG which 
could be formally written as:

\begin{eqnarray}
\varepsilon_c^{(k+1)}&= &f\left(\varepsilon_c^{(k)},
\delta
\varepsilon^{(k)}; \rho^{(k)},
{\bf P}^{(k)}\right)\nonumber\\ &&\\
\delta \varepsilon^{(k+1)}&=
&g\left(\varepsilon_c^{(k)},\delta\varepsilon^{(k)};
\rho^{(k)}, {\bf
P}^{(k)}\right)\nonumber
\end{eqnarray}

\noindent where $\rho^{(k)}$ 
(density of critical sites) and ${\bf
P}^{(k)}\equiv(p_1^{(k)},p_2^{(k)},p_3^{(k)},
p_4^{(k)})$ ($p_n$ is the
probability that in a toppling the energy 
is distributed exactly to
$n$ nearest neighbours) are the parameters 
characterizing the state at
the generic scale $k$.  In principle we have 
to couple these two
equations to those that determine the 
renormalization of $\rho^{(k)}$
and ${\bf P}^{(k)}$; for sake of simplicity, 
having extended the
parameters space, we assume that the flow 
of the new parameters is
orthogonal to the space defined by the old ones. 
We then accordingly
fix $\rho^{(k)}$ and ${\bf P}^{(k)}$ to their 
fixed point values
$\rho^*$ and ${\bf P}^*$ as computed in ref.[9].
Their
values are $\rho^*=0.468$ and ${\bf P}^*= 
(0.240,0.442,0.261,0.057)$. The
validity of this approximation will be checked by numerical
simulations. Our transformation will have the form:

\begin{eqnarray}
\varepsilon^{(k+1)}&=&f\left(\varepsilon_c^{(k)},
\delta\varepsilon^{(k)};\rho^*,
{\bf P}^*\right)\nonumber\\
&&\\
\delta\varepsilon^{(k+1)}&=&g\left(\varepsilon_c^{(k)},
\delta
\varepsilon^{(k)};\rho^*,
{\bf P}^*\right)\nonumber
\end{eqnarray}

\begin{figure}
\protect
\centerline{\psfig{figure=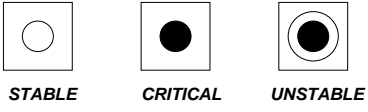,width=5cm,angle=0}}
\caption{\small Critical, stable and unstable sites.}
\label{scu}
\end{figure}

\begin{figure}
\protect
\centerline{\psfig{figure=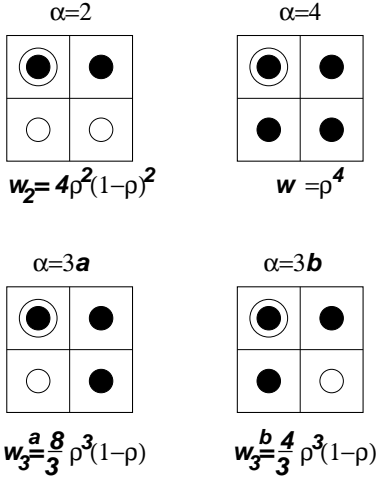,width=5cm}}
\caption{\small The $\alpha$ configurations 
with relative statistical 
weights.}
\label{statw}
\end{figure}

\noindent The symbols used are shown in Fig.~\ref{scu} and
~\ref{statw}.

\begin{figure}
\protect
\centerline{\psfig{figure=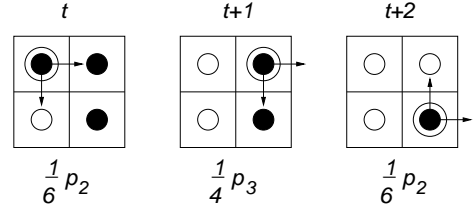,height=3cm,angle=0}}
\caption{One of the possible time evolutions 
starting from an 
$\alpha=3$ configuration and transferring 
energy outside 
of the cell; for every step the relative 
probability 
is indicated.}
\label{fig1}
\end{figure}

Let us consider now a single path as an example 
of the transformation
we want to obtain. We have a $2\times 2$ cell 
characterized by a label
$\alpha$ specifying the number of critical sites. 
The $\alpha=1$
configuration is not considered because of a 
\emph{spanning rule} that
imposes the exclusion of those processes 
that do not relax within the
cell before they transfer energy outside.  
Every configuration
$\alpha$ has a mean height 
$\langle\varepsilon_\alpha\rangle$ that we
calculate in the next section. 
One of the possible evolution from the
configuration $\alpha=3a$ state is shown in 
Fig.~\ref{fig1}.

This means that at the scale $k+1$ the 
mean height of the considered
cell is critical with probability \be \pi\equiv
w_{\alpha=3}^{(a)}(\rho^*)\cdot \frac16 p_2^*\cdot\frac14
p_3^*\cdot\frac16 p_2^* \ee As a matter of fact, 
the cell has a
probability $\pi$ to be a critical site at 
the scale $k+1$. The
complete renormalization flow is therefore 
a sum over all the
configuration and over all the possible paths 
of terms like
$\pi\cdot\langle\varepsilon_{\alpha=3}^{(k)}
\rangle$.  In the same
process we observe that with probability 
$\pi$ the quantity given
outside of the cell, 
$\delta\varepsilon^{(k+1)}$, is $3/2\cdot
\delta\varepsilon^{(k)}$. $3$ flows of 
energy $\delta
\varepsilon^{(k)}$ are lost outside the cell 
passing through $2$
sides. It means that for this transformation 
we have to distinguish
paths transferring the same energy through a 
different number of
sides.  The general transformations are
\begin{eqnarray}
\varepsilon_c^{(k+1)}&=&\sum_\alpha 
w_\alpha(\rho^*)\,\left[\langle
\varepsilon_\alpha^{(k)}\rangle\cdot\sum_i 
P_{i,\alpha}({\bf
P}^*)\right]\nonumber\\ &&\\ \delta 
\varepsilon^{(k+1)}&=&\sum_\alpha
w_\alpha(\rho^*) \,\left[\sum_i 
P_{i,\alpha}^{n,m}({\bf P}^*)\, \frac
n m\cdot\delta \varepsilon^{(k)}\right]\nonumber
\end{eqnarray}
where $P_{i,\alpha}$ are the probabilities 
relative to time evolutions
that transfer energy outside of the cell, 
while $P_{i,\alpha}^{n,m}$
refers to a process that let $n$ quanta going 
through $m$ sides.

We compute now the average energy per site in 
all the configurations.
We generalize the simple mean field argument 
described
in ref.[11],
making the simpler non trivial assumption for the
probability distribution of the energy:
$P(\varepsilon)=a+b\,\varepsilon$. The 
normalization \be
\int_0^{\varepsilon_c}d\varepsilon\,P(\varepsilon)=1 
\ee implies that
$b=2(1-a\,\varepsilon_c)/{\varepsilon_c^2}$.
 
The average energy is then expressed in term 
of the parameter $a$: \be
\langle\varepsilon \rangle\equiv\int_0^
{\varepsilon_c}d\varepsilon
\,\varepsilon\,P(\varepsilon)= \frac 23\,
\varepsilon_c-\frac
16\,a\varepsilon_c^2\;. \ee Following ref.[11]
we find that
$a\simeq 0.277/\varepsilon_c$.  
The average energy of stable and
critical sites is then:
\begin{eqnarray}
\langle \varepsilon\rangle_{stable} &\equiv&\frac
{\int\limits_0^{\varepsilon_c- 
\delta\varepsilon }d\varepsilon\,
\varepsilon\,P(\varepsilon)}{\int\limits_0^
{\varepsilon_c -\delta
\varepsilon}d\varepsilon\,P(\varepsilon)}= \frac
{\int\limits_0^{\varepsilon_c-\delta \varepsilon}
d\varepsilon\,\left(a\,\varepsilon+ \frac
2{\varepsilon_c^2}\,\varepsilon^2\,(1-a\, 
\varepsilon_c)\right)}
{\int\limits_0^{\varepsilon_c-\delta \varepsilon}
d\varepsilon\,\left(a+\frac 2{\varepsilon_c^2}\,
\varepsilon
\,(1-a\,\varepsilon_c)\right)} \nonumber\\
&=&\frac{\frac 23-\frac16\,a\,
\varepsilon_c-\frac{\delta \varepsilon}
{\varepsilon_c}\left(\frac43-\frac56\,a\,
\varepsilon_c\right)+
O(\delta \varepsilon)^2}{\frac 1{\varepsilon_c}
\left[ 1-\frac{\delta
\varepsilon}{\varepsilon_c} \left(1-a\,
\varepsilon_c\right)\right]}
\nonumber\\ 
&=& \langle \varepsilon\rangle-\frac23\delta 
\varepsilon\left(
1-\frac14a^2\varepsilon_c^2\right)+O(\delta 
\varepsilon)^2
\end{eqnarray}

\begin{eqnarray}
\langle \varepsilon\rangle_{critical}&\equiv&
\frac{\int\limits_{\varepsilon_c-\delta \varepsilon}^
{\varepsilon_c}d\varepsilon\,
\varepsilon\,P(\varepsilon)}
{\int\limits_{\varepsilon_c- \delta
\varepsilon}^{\varepsilon_c}d\varepsilon\,P(\varepsilon)}= 
\frac
{\int\limits_{\varepsilon_c-\delta
\varepsilon}^{\varepsilon_c}d\varepsilon\left
(a\varepsilon+\frac 2{\varepsilon_c^2}\varepsilon^2 
(1-a\varepsilon_c)\right)}
{\int\limits_{\varepsilon_c-\delta \varepsilon}
^{\varepsilon_c}d\varepsilon\left(a+\frac
2{\varepsilon_c^2}\varepsilon(1-a\varepsilon_c)\right)}
\nonumber\\
&=&\frac{\varepsilon_c(2-a\,\varepsilon_c)-
\delta \varepsilon
\left(2-\frac72\,a\,\varepsilon_c\right)}
{(2-a\,\varepsilon_c)\left[1-\frac{\delta \varepsilon}
{\varepsilon_c}\frac{1-a\,\varepsilon_c}{2-a\,
\varepsilon_c}\right]}
\nonumber\\ 
&=& \varepsilon_c-\delta
\varepsilon\,\frac{1-\frac52\,a\,\varepsilon_c}
{2-a\,\varepsilon_c}+O(\delta \varepsilon)^2
\end{eqnarray}
If we put $a=0.277/\varepsilon_c$ in these 
expressions we obtain 
\be
\langle \varepsilon\rangle_{stable}=0.6205\,\varepsilon_c-
0.539\,\delta \varepsilon \ee \be \langle
\varepsilon\rangle_{critical}=\varepsilon_c- 0.1785\,\delta
\varepsilon \ee 
The average energy per site for different
configurations can now be calculated as 
\be \langle
\varepsilon_{\alpha=2}\rangle =\frac14\left[2\, \langle
\varepsilon\rangle_{critical}+ 2\,\langle
\varepsilon\rangle_{stable}\right]=0.810\, \varepsilon_c-0.359\,\delta
\varepsilon \ee \be \langle \varepsilon_{\alpha=3}\rangle=
\frac14\left[3\, \langle 
\varepsilon\rangle_{critical}+ \langle
\varepsilon\rangle_{stable}\right]=0.905\,
\varepsilon_c-0.269\,\delta
\varepsilon \ee \be \langle 
\varepsilon_{\alpha=4}\rangle = \langle
\varepsilon\rangle_{critical}= \varepsilon_c-0.1785\,\delta
\varepsilon \ee
Equations for the renormalization flow have the form
\be \left\{
\begin{array}{rcl}
\varepsilon_c^{(k+1)}&=&a\,\varepsilon_c^{(k)}-b\,
\delta \varepsilon^{(k)}\\
\delta\varepsilon^{(k+1)}&=&c\,\delta\varepsilon^{(k)}
\end{array}
\right.
\ee
where the coefficients $a,b$ and $c$ obviously 
depend on the 
fixed point parameters. If we divide the two 
equations, recalling 
the definition of the rigidity, we obtain
\be
r^{(k+1)}=\frac ac\,r^{(k)}-\frac bc\;.
\ee
The fixed point of the rigidity 
corresponds to setting $r^{(k+1)}=
r^{(k)}=r^*$:
\be
r^*=\frac b{a-c}\;.
\ee
Next section deals with the calculation of the coefficients 
$a,b$ and $c$.

\section{Analytical results}
To calculate the coefficients $a,b$ and $c$ 
we have to distinguish 
between $4$ different kind of configurations, 
each one with the relative 
weights $w_\alpha(\rho)$ as calculated in ref.[9].

$\bullet$ {\bf $\alpha=2$}

\begin{figure}
\protect
\centerline{\psfig{figure=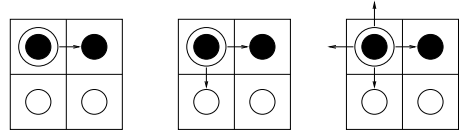,width=6cm,angle=270}}
\caption{\small Time $t$: the probability relative 
to the paths in figure is 
$\frac 1 4 p_1+\frac 1 2 p_2+\frac 3 4 p_3 +p_4$.}
\label{3a2t1}
\end{figure}

\begin{figure}
\centerline{\psfig{figure=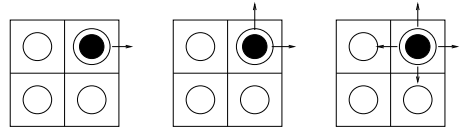,width=6cm}}
\caption{\small Time $t+1$: the probability associated with 
the paths in figure is 
$\frac 1 2 p_1+\frac 56 p_2+p_3 +p_4$.}
\label{3a2t2}
\end{figure}

We consider at time $t$ one of the toppling represented in
Fig.~\ref{3a2t1} and possible developments at time $t+1$ in
Fig.~\ref{3a2t2}.

All the paths described transfer energy outside 
the cell, the associated probabilities 
are given in captions. 
Configuration $\alpha=2$ then contributes with the term
\[
\varepsilon_c^{(k+1)}=
w_{\alpha=2}(\rho)\cdot\langle 
\varepsilon_{\alpha=2}^{(k)}\rangle
\cdot\left(\frac14p_1+\frac12p_2+
\frac34p_3+p_4\right)\times
\]
\be
\times
\left(\frac12p_1+\frac56p_2+p_3+p_4
\right)+\ldots
\ee
to the renormalization of the critical energy. 
Moreover, each of the 
described paths transfer only a quantum of 
energy outside of the cell, so
\[
\frac{\delta \varepsilon^{(k+1)}}
{\delta \varepsilon^{(k)}}=w_{\alpha=2}(\rho)\cdot
\left(\frac 1 4 p_1+\frac 1 2 p_2+
\frac 3 4 p_3 +p_4\right)
\times\]
\be\times
\left(
\frac 1 2 p_1+\frac 56 p_2+p_3 +p_4
\right)+\ldots
\ee

The details of the calculations regarding paths from 
configurations $\alpha=3a,\,3b$ and $4$ are given in the 
appendix. 

The final result is obtained by summing 
all these contributes that we
have explicitly calculated with the respective weights.  
We find for
the coefficients the following values: 
\be a=0.26\pm0.03\quad
b=0.10\pm0.02\quad c=0.26\pm0.05 \ee 
We finally obtain that $a$ and
$c$ are equal, even though the errors 
associated are relatively
large. These errors are related to 
the parameters of the fixed point.
Since they have only $3$ significant digits, 
that results an error of
order $10^{-3}$.

We then find that the fixed point spontaneously 
reached by a sandpile
model corresponds to an infinite value of 
the rigidity. It means that
every finite value for the 
microscopic dynamics renormalizes to an
infinite one on a large scale. 

The difference with a diffusive system
is now clear: in the diffusive case a 
particle added from outside can
always goes out of the system from the 
boundaries, while in a SOC
system this is avoided by the value of 
the rigidity as bigger as
observed from a large scale.

\section{Numerical Results}
We present here the results of 
simulations made on BTW~\cite{btw} and
Zhang~\cite{zh} models.  We have measured 
for these systems the
critical energy and the average flow 
transferred from the boundaries
at a generic scale $k$. Critical 
energy is defined as follows: we
consider an external perturbation on a site 
in a coarse grained cell
($k$-cell) of size $l^{(k)}\times l^{(k)}$ 
and we look at the
avalanches starting from this coarse-grained cell. 
If the avalanche
transfers energy outside the given $k$-cell
 we compute the average
energy $\varepsilon_c^{(k)}$ of that $k$-cell 
before the toppling; as
a matter of fact, that $k$-cell is critical 
with a probability given
by the frequency of the avalanches transferring 
energy outside the
$k$-cell itself. Moreover, we define the energy 
transferred $\delta
\varepsilon^{(k)}$ at the scale $l^{(k)}$ 
as the number of grains
going outside the $k$-cell divided by the 
number of boundaries crossed
and by the length $l^{(k)}$ 
(this is the trivial scaling).

\begin{figure}
\protect
\centerline{\psfig{figure=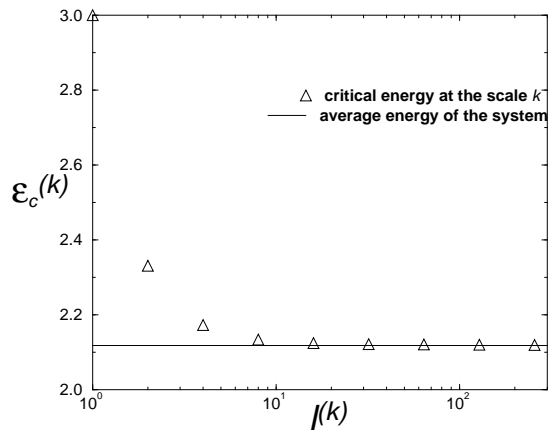,width=7cm,angle=0}}
\caption{\small The behaviour of the critical 
energy $\varepsilon_c^{(k)}$ at
the scale $k$ in the BTW sandpile model.}
\label{zc}
\end{figure}

\begin{figure}
\protect
\centerline{\psfig{figure=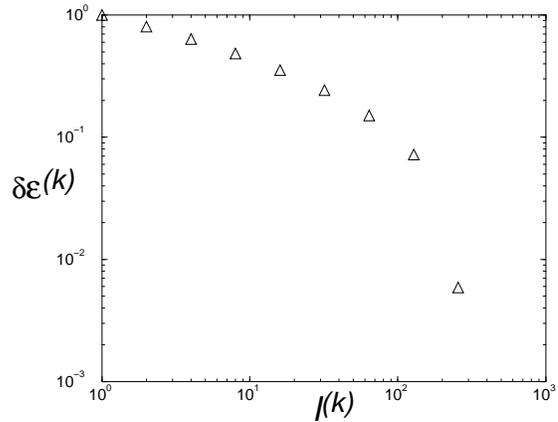,width=7.2cm,angle=0}}
\caption{\small Numerical 
renormalization of $\delta\varepsilon$ 
in the BTW.}
\label{dz}
\end{figure}

\begin{figure}
\protect
\centerline{\psfig{figure=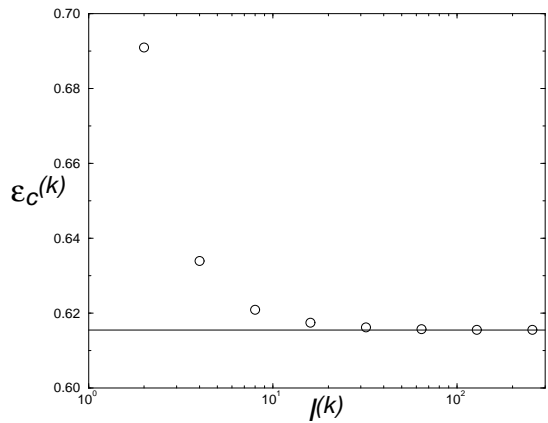,width=7cm,angle=270}}
\caption{\small  The behaviour of the critical 
energy at a generic
scale $k$ for the Zhang model.}
\label{zczhang}
\end{figure}

\begin{figure}
\protect
\centerline{\psfig{figure=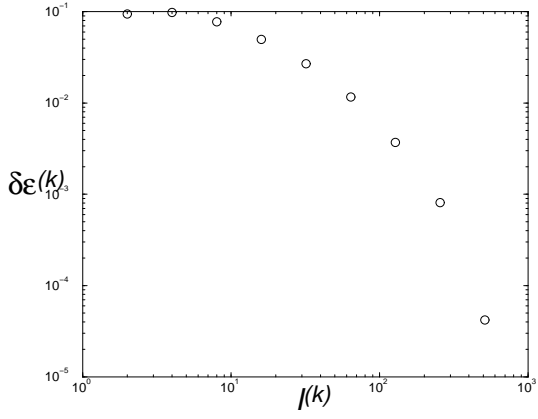,width=7cm,angle=270}}
\caption{\small 
Numerical renormalization of 
$\delta\varepsilon$ in the 
Zhang model.}
\label{dzzhang}
\end{figure}

If we look at numerical results (Fig.~\ref{zc}) 
we note that the
critical energy at large scale is the average 
energy~\cite{pr94} of
the system and this is not surprising 
since at the scale of the system
size this is exactly what we compute. 
As for the behaviour of
$\delta\varepsilon$ (Fig.~\ref{dz}) 
one gets that in the limit
$k\rightarrow\infty$ it goes to zero. 
This result confirms the idea
that the fixed point value for the 
rigidity is infinite, signalling a clear
difference with respect to usual diffusive systems.  
The same qualitative results are obtained for the Zhang
model~\cite{zh} (See Fig.~\ref{zczhang} and ~\ref{dzzhang}).

\section{Conclusions}

In summary we have investigated, 
both in the framework of the
Dynamically Driven Renormalizaton 
Group scheme and by numerical
simulations, the role of the 
so-called local rigidity in sandpile
models. The local rigidity is defined 
as the ratio between the
critical energy and the amplitude 
of the external perturbation, both
at the microscopic scale.  It turns out that, under the
renormalization flow equations of the 
DDRG, the local rigidity
renormalizes to an infinite fixed 
point value both for the BTW model
and the Zhang model, while for a 
typical diffusive system one would
expect a vanishing value. 
Numerical simulations of coarse-grained
sandpile models confirm this picture.


Authors acknowledge EU contract FMRXCT980183 and
G.C. acknowledge
FET Open project COSIN
IST-2001-33555.

\section{Appendix}

$\bullet$ {\bf $\alpha=3a$}

\begin{center}
\begin{figure}
\centerline{\psfig{figure=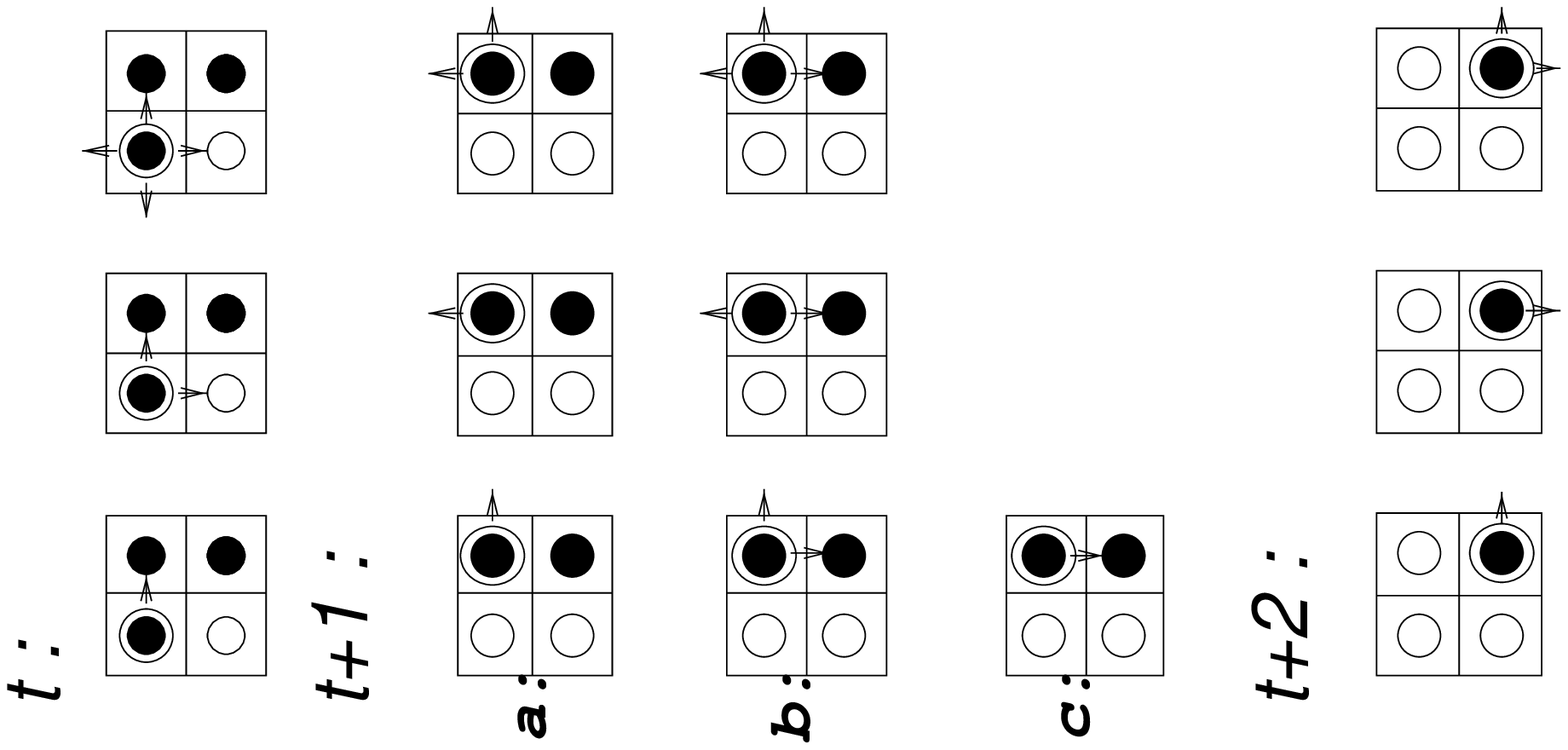,width=17cm,angle=270}}
\caption{\small Time $t$: the probability relative 
to the paths in figure is 
$\frac 1 4 p_1+\frac 1 2 p_2+\frac 3 4 p_3 +p_4$.}
\label{3a3at}
\end{figure}
\end{center}

In Fig.~\ref{3a3at} we show all 
the possible paths starting from the 
perturbation of the critical site. 

The probability relative to the first 
event at time $t$ is the same as before, 
$\frac14p_1+\frac12p_2+\frac34p_3+p_4$. 
At time $t+1$ we have 
$a$) $\frac 14 p_1+\frac16 p_2$ for the first two 
events in figure, $\frac16 p_2+\frac14 p_3$ 
for the third 
(they are distinguishable only with 
respect to renormalization 
of $\delta \varepsilon$);
$b$)  $\frac16p_2+\frac14p_3$ for the first two events, 
$\frac14p_3+p_4$ for the third;
$c$) $\frac14p_1+\frac16p_2$.

At time $t+2$ we have 
$\frac14p_1+\frac13p_2+\frac14p_3$ for the first 
two and $\frac16p_2+\frac12p_3+p_4$ 
for the third; in the renormalization 
of $\varepsilon_c$ we have to consider only the sum 
$\frac12p_1+\frac56p_2+p_3+p_4$. So we can write
\[
\varepsilon_c^{(k+1)}=\ldots+w_{\alpha=3}^{(a)}(\rho)\cdot
\langle \varepsilon_{\alpha=3}^{(k)}
\rangle\cdot\left(\frac14p_1+\frac12p_2+\frac34p_3+p_4
\right)\times
\]
\be
\times
\left[\left(\frac12p_1+\frac56p_2+p_3+p_4\right)
\left(1+\left(\frac14p_1+
\frac16p_2\right)\right)\right]+\ldots
\ee
and
\[
\frac{\delta \varepsilon^{(k+1)}}
{\delta \varepsilon^{(k)}}=
\ldots+w_{\alpha=3}^{(a)}(\rho)\cdot
\left(\frac14p_1+\frac12p_2+\frac34p_3+p_4\right)
\times\]
\[\times
\left[\left(\frac12p_1+\frac12p_2+\frac14p_3\right)\right.
+\left(\frac16p_2+\frac14p_3\right)
\times\]
\[\times
\left(\left(
\frac14p_1+\frac13p_2+\frac14p_3\right)\left(2+1\right)\right.
+\left.\left(\frac16p_2+\frac12p_3+p_4\right)\frac32\right)+
\]
\[
+\left(\frac16p_2+\frac14p_3\right)\left(\frac12p_1+
\frac56p_2+p_3+p_4\right)
+\left(\frac14p_3+p_4\right)
\]
\[
\left(\left(\frac14p_1+\frac13p_2+
\frac14p_3\right)\left(\frac32+1\right)+
\left(\frac16p_2+
\frac12p_3+p_4\right)\frac43\right)\times
\]
\be\times
\left.\left(\frac14p_1+\frac16p_2\right)\left(
\frac12p_1+\frac56p_2+p_3+p_4\right)\right]+\ldots
\ee
for the contributions coming from the $\alpha=3a$ 
configuration.              
 
$\bullet$ {\bf $\alpha=3b$}

\begin{figure}
\protect
\centerline{\psfig{figure=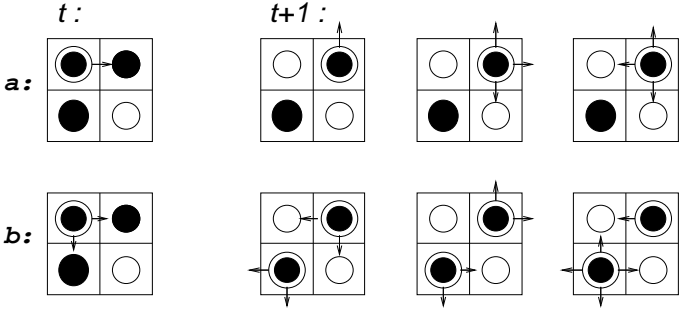,width=9cm,angle=0}}
\caption{\small Evolutions for the $\alpha=3b$ 
configuration.}
\label{3a3bt}
\end{figure}

We can compute the probabilities of the shown in 
Fig.~\ref{3a3bt}.
 
At time $t$ we have to distinguish 
between two kinds of events: the 
toppling on a single site (this term 
is considered two times to include 
the toppling on the other site) and 
the one on two sites. The statistical 
weights are $\left(\frac14p_1+\frac13p_2+
\frac14p_3\right)$ and 
$\left(\frac16p_2+\frac12p_3+p_4\right)$ 
respectively. At time $t+1$ 
both the events leave one flow going 
outside the cell, so we have:
\[
\varepsilon_c^{(k+1)}=\ldots+w_{\alpha=3}^{(b)}(\rho)\cdot
\langle \varepsilon_{\alpha=3}^{(k)}\rangle
\times\]
\[\times
\left[2\cdot\left(\frac14p_1+\frac12p_2+
\frac34p_3+p_4\right)
\cdot\left(\frac12p_1+\frac56+p_3+p_4\right)+\right.
\]
\be
+\left.\left(\frac16p_2+\frac12p_3+p_4\right)\cdot
\left(1-\left(\frac12p_1+\frac56p_2\right)^2\right)
\right]+\ldots
\ee
\[
\frac{\delta \varepsilon^{(k+1)}}
{\delta \varepsilon^{(k)}}
=\ldots+w_{\alpha=3}^{(b)}(\rho)\times
\]
\[\times
\left[2\cdot\left(\frac14p_1+\frac12p_2+\frac34p_3+
p_4\right)
\cdot\left(\frac12p_1+\frac56+p_3+p_4\right)\right.+
\]
\be
+\left.\left(\frac16p_2+\frac12p_3+p_4\right)\cdot
\left(1-\left(\frac12p_1+\frac56p_2\right)^2\right)
\right]+\ldots
\ee

$\bullet$ {\bf $\alpha=4$} 

\begin{figure}
\protect
\centerline{\psfig{figure=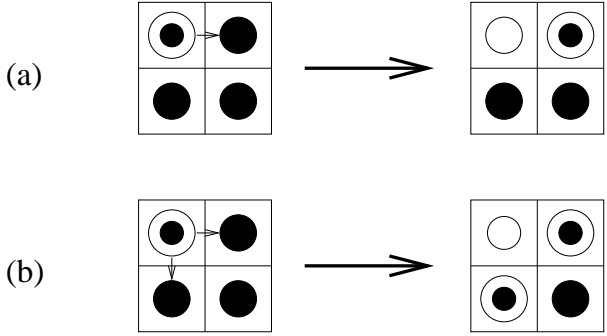,width=8cm,angle=0}}
\caption{\small Events $a$ and $b$ for the $\alpha=4$ 
configuration.}
\label{3a4t0}
\end{figure}

\begin{figure}
\protect
\centerline{\psfig{figure=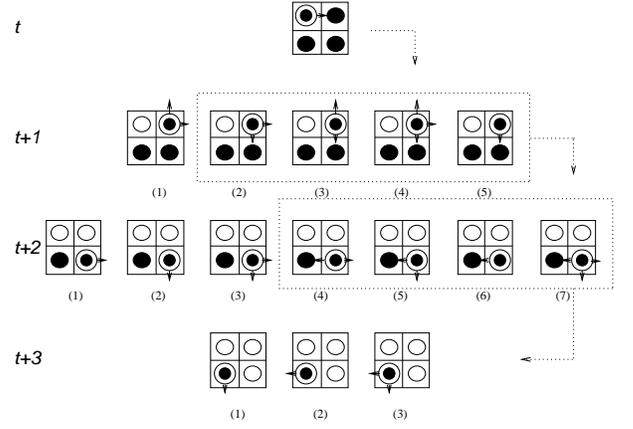,width=8cm,angle=0}}
\caption{\small Possible paths from event $a$.}
\label{3a4t}
\end{figure}

In a cell with all critical sites the number of possible 
time evolutions grows enormously. 
In Fig.~\ref{3a4t0} we show 
the first two possible events, with probabilities 
$p(a)=\frac14p_1+\frac13p_2+\frac14p_3$ and
$p(b)=\frac16p_2+\frac12p_3+p_4$. The paths coming 
from $a$ are shown in Fig.~\ref{3a4t} 
and the relatives probabilities 
in Table~\ref{tab}.

\begin{table}[h]
\centering
\begin{tabular}{|c|c|c|c|}
\hline
&$t+1$&$t+2$&$t+3$\\
\hline
(1)&$\frac12p_1+\frac12p_2+\frac14p_3$&
$\frac14p_1+\frac16p_2$&
$\frac14p_1+\frac13p_2+\frac14p_3$\\
\hline
(2)&$\frac16p_2+\frac14p_3$&
$\frac14p_1+\frac16p_2$&
$\frac14p_1+\frac13p_2+\frac14p_3$\\
\hline
(3)&$\frac16p_2+\frac14p_3$&$\frac16p_2+\frac14p_3$&
$\frac16p_2+\frac12p_3+p_4$\\
\hline
(4)&$\frac14p_3+p_4$&$\frac16p_2+\frac14p_3$&///\\
\hline
(5)&$\frac14p_1+\frac16p_2$&$\frac16p_2+\frac14p_3$&///\\
\hline
(6)&///&$\frac14p_3+p_4$&///\\
\hline
(7)&///&$\frac14p_1+\frac16p_2$&///\\
\hline
\end{tabular}
\caption{\small Probabilities relative to the events in
Fig.~\ref{3a4t}.}
\label{tab}
\end{table}

Now we have to compute $224$ paths from the $b$ event.  
The final result, avoiding details, for the renormalization 
of the critical height is 
\[
\varepsilon_c^{(k+1)}=
\ldots + w_{\alpha=4}(\rho)\cdot\left\{
2\cdot\left(\frac14p_1+\frac13p_2+\frac14p_3\right)
\right.\times\]
\[\times
\left[\left(\frac12p_1+\frac56p_2+p_3+p_4\right)\cdot
\right.
\left.\left(1+\left(\frac14p_1+\frac16p_2\right)+
\right.\right.
\]
\[+\left.\left.
\left(\frac14p_1+\frac16p_2\right)^2\right)\right]
+\left(\frac16p_2+\frac12p_3+p_4\right)\cdot
\left[1-\left(\frac14p_1\right)^2\right.
+\]
\be\left.\left.
-\left(\frac14p_1+\frac16p_2\right)
\cdot\left(\frac34p_1+\frac16p_2\right)\cdot
\left(\frac12p_1+\frac16p_2\right)\right]\right\}
\ee
If we introduce the quantities
\begin{eqnarray}
\pi_1&\equiv&\frac14p_1+\frac12p_2+\frac34p_3+p_4\nonumber\\
\pi_2&\equiv&\frac12p_1+\frac56p_2+p_3+p_4\nonumber\\
\pi_3&\equiv&\frac12p_1+\frac12p_2+\frac14p_3\nonumber\\
\pi_4&\equiv&\frac16p_2+\frac14p_3\nonumber\\
\pi_5&\equiv&\frac14p_1+\frac13p_2+\frac14p_3\nonumber\\
\pi_6&\equiv&\frac16p_2+\frac12p_3+p_4\nonumber\\
\pi_7&\equiv&\frac14p_1+\frac16p_2\nonumber\\
\pi_8&\equiv&\frac12p_1+\frac16p_2\nonumber\\
\pi_9&\equiv&\frac14p_3+p_4\nonumber\\
\pi_{10}&\equiv&\frac34p_1+\frac12p_2+\frac14p_3\nonumber
\end{eqnarray}
then we can write down the renormalization of $\delta\varepsilon$ as 
\begin{eqnarray}
\frac{\delta E^{(k+1)}}{\delta E^{(k)}}
&&\hspace{-0.3cm}
=\ldots+w_{\alpha=4}(\rho)\cdot
\left\{
2\,\pi_5\,\left[\pi_3+\right.\right.
\nonumber\\
&&\hspace{-1.7cm}
\pi_4\,\left(2\,\pi_7+
\pi_4\,\left(1+\pi_2+\frac52\pi_5+\frac43\pi_6
\right)\right)+
\nonumber\\
&&\hspace{-1.7cm}
+\pi_9\,\left(\frac43\pi_5+\frac54\pi_6\right)+
\pi_7\,\pi_2+\pi_7\,(2+1)+
\nonumber\\
&&\hspace{-1.7cm}
+\pi_4\,\left(
\frac32+\left(\frac32+\frac32\right)\pi_5+\pi_6\right)+
\pi_4\,\left(\left(\frac32+1\right)\pi_5+\frac43\pi_6\right)+
\nonumber\\
&&\hspace{-1.7cm}
\left.
+\pi_9\,\left(\left(2+\frac43\right)\pi_5+\frac53\pi_6\right)+
\pi_7\,\pi_2\right)+\pi_9\times
\nonumber\\
&&\hspace{-1.7cm}
\times\left(\left(\frac32+1\right)\pi_7+\right.
+\pi_4\,\left(\frac43+\frac83\pi_5+\frac54\pi_6\right)+
\nonumber\\
&&\hspace{-1.7cm}
\pi_4\,\left(\frac73\pi_5+\frac54\pi_6\right)+
+\pi_9\left(\left(\frac53+\frac54\right)\pi_5+\right.
+\left.\left.\frac32\pi_6\right)+\right.
\nonumber\\
&&\hspace{-1.7cm}
\left.
+\pi_7\,\pi_2\right)+\pi_7\,\left(\pi_3+\pi_4\,
\left(\pi_2+3\pi_5\frac32\pi_6\right)\right.+
\nonumber\\
&&\hspace{-1.7cm}
+\left.\left.\pi_9\,\left(\frac52\pi_5+\frac43\pi_6\right)+
\pi_7\,\pi_2\right)\right]+\pi_6\times
\nonumber\\
&&\hspace{-1.7cm}
\left[
\left(\pi_{10}^2-\frac1{16}p_1^2\right)+\right.
+2\,\left(\pi_7\,\pi_5+\frac14p_1\,\pi_4\right)\,
\left(5\pi_5+\frac52\pi_6\right)+
\nonumber\\
&&\hspace{-1.7cm}
2\cdot
\left(\pi_7\,\pi_6+
\frac14p_1\,\pi_9\right)\,\left(\frac52\pi_5+\frac43\pi_6\right)+
\nonumber\\
&&\hspace{-1.7cm}
+\left(2\,\pi_7\,\pi_4+\pi_4^2\right)
\left(10\pi_5+\frac{17}3\pi_6\right)+
\left(2\pi_7\,\frac14p_1+\pi_7^2\right)\,\pi_2+
\nonumber\\
&&\hspace{-1.7cm}
+\left.\left.2\left(\pi_4^2+\pi_9\,\pi_5\right)
\left(5\pi_5+\frac{35}{12}\pi_6\right)+
\left(2\,\pi_9\,\pi_4+\pi_9^2\right)\times\right.\right.
\nonumber\\
&&\hspace{-1.7cm}
\times\left.\left.
\left(\frac52\pi_5+\frac32\pi_6\right)\right]\right\}
\end{eqnarray}



\begin{references}
\vspace*{-0.3in}

\bibitem{btw} P.~Bak, C.~Tang, K.~Wiesenfeld \e{Phys.~Rev.~Lett.} 
\bb{4}, 381 (1987); \e{Phys.~Rev.~A} \bb{38}, 364 (1988)

\bibitem{jen} H.J.~Jensen, \e{Self-Organized Criticality},  
Cambridge University Press (1998)

\bibitem{review_braz} R.~Dickman, M.A.~Mu\~noz, A.~Vespignani and
S. Zapperi, {\em Paths to Self-Organized Criticality}, Short
review in {\em Brazilian Journal of Physics}, {\bf 30}, 27 (2000)
and references therein.

\bibitem{vz97} A.~Vespignani, S.~Zapperi \e{Phys.~Rev.~Lett.} \bb{78}, 
4793 (1997).


\bibitem{vzl96} D.~Sornette, A.~Johansen, I.~Dornic \e{J.~Phys.~I} 
(France) \bb{5}, 325 (1995); A.~Vespignani, S.~Zapperi, 
V.~Loreto \e{Phys.~Rev.~Lett.} \bb{77}, 4560 (1996); 
\e{J.~Stat.~Phys.} \bb{88}, 47 (1997)

\bibitem{verg} M.~Vergeles \e{Phys.~Rev.~Lett.} \bb{75}, 1969 (1995)

\bibitem{gcp} A.~Vespignani, S.~Zapperi \e{Phys.~Rev.~E} \bb{57},
6345 (1998), A.~Gabrielli, G.~Caldarelli, L.~Pietronero 
\e{Phys.~Rev.~E \bb{62}, 7638} (2000) 

\bibitem{caf} R.~Cafiero, V.~Loreto, L.~Pietronero, A.~Vespignani, 
S.~Zapperi \e{Europhys.~Lett.} \bb{29}, 111 (1994)

\bibitem{pvz94} L.~Pietronero, A.~Vespignani, S.~Zapperi 
\e{Phys.~Rev.~Lett.} \bb{72}, 1690 (1994); A.~Vespignani, S.~Zapperi, 
L.~Pietronero \e{Phys.~Rev.~E} \bb{51} 1711 (1995)

\bibitem{ddrg} A. Vespignani, S. Zapperi and V. Loreto, {\em
Phys. Rev. Lett.} {\bf 77}, 4560 (1996); A. Vespignani, S. Zapperi and
V. Loreto, {\em J. of Stat. Phys.} {\bf 88}, 47 (1997).

\bibitem{pitazh88} L.~Pietronero, P.~Tartaglia, Y.-C.~Zhang \e{Physica
A} \bb{173}, 129 (1991)

\bibitem{pr94} V.B.~Priezzhev \e{J.~Stat.~Phys.} \bb{74}, 955 (1994)

\bibitem{zh} Y.-C.~Zhang \e{Phys.~Rev.~Lett.} \bb{63}, 470 (1989)

\end{references}
\end{document}